\documentclass[twocolumn,nofootinbib,notitlepage]{revtex4-1}
\usepackage{epsfig,amsmath,slashed}
\usepackage{tensor}

\newcommand{\di}{{\rm d}}

\newcommand{\e}{{\rm e}}

\newcommand{\be}{\begin{equation}}
\newcommand{\ee}{\end{equation}}                                                                               
\newcommand{\bea}{\begin{eqnarray}}
\newcommand{\eea}{\end{eqnarray}}                  
\newcommand{\snn}{\sqrt{s_{\rm NN}}}

\begin{document}

\title{Collective longitudinal polarization in relativistic heavy-ion collisions
at very high energy} 

\author{F.~Becattini}
\author{Iu.~Karpenko}
\affiliation{Universit\`a di Firenze and INFN Sezione di Firenze, Via G. Sansone 1, I-50019 
Sesto Fiorentino (Firenze), Italy} 

\begin{abstract}
We study the polarization of particles in relativistic heavy-ion collisions 
at very high energy along the beam direction within a relativistic hydrodynamic
framework. We show that this component of the polarization decreases much slower 
with center-of-mass energy compared to the transverse component, even in the 
ideal longitudinal boost-invariant scenario with non-fluctuating initial state, 
and that it can be measured 
by taking advantage of its quadrupole structure in the transverse momentum plane. 
In the ideal longitudinal boost-invariant scenario, the polarization is proportional to 
the gradient of temperature at the hadronization and its measurement can provide 
important information about the cooling rate of the Quark Gluon Plasma around 
the critical temperature.
\end{abstract}

\maketitle

Global polarization of hadrons produced in relativistic heavy-ion collisions has
been recently observed by the STAR experiment over a center-of-mass energy range
between 7.7 and 200 GeV \cite{STAR:2017ckg,Abelev:2007zk}. This finding confirms 
early proposals \cite{Liang:2004ph, Liang:2004xn}, later predictions based on local 
thermodynamic equilibrium of spin degrees of freedom \cite{Becattini:2013fla,Becattini:2013vja} 
which provides a relation between polarization and relativistic vorticity, and it 
agrees quantitatively with the 
hydrodynamic model calculations \cite{Becattini:2016gvu,Karpenko:2016jyx,Xie:2017upb} 
to a very good degree of accuracy. In the hydrodynamic framework, the distinctive 
feature of polarization is its proportionality to the gradients of the combined 
temperature and velocity fields (see eq.~(\ref{mainf})), so that its measurement 
is a stringent test of the hydrodynamic picture which is distinct and complementary 
to momentum spectra. 

The experimental efforts, thus far, focused on the search of the average global 
polarization of $\Lambda$ hyperons along the direction of the angular momentum of 
the plasma. This measurement requires the identification of the reaction plane in 
peripheral collision as well as its orientation, that is the direction of the total 
angular momentum vector ${\bf J}$. 
The average global polarization along ${\bf J}$ is also found to decrease rapidly 
as a function of center-of-mass energy \cite{STAR:2017ckg}, from few percent at 
$\snn = {\cal O}(10)$ GeV to few permille at $\snn = {\cal O}(100)$ GeV, in agreement 
with calculations based on the hydrodynamic model \cite{Karpenko:2016jyx,Xie:2017upb,
Li:2017dan} as well with hybrid approaches \cite{Li:2017slc,Sun:2017xhx}. In the TeV 
energy range, at the LHC, the global polarization along ${\bf J}$ is not seen 
\cite{alice} as it is most likely beyond experimental sensitivity. 

It is of course desirable to check more - possibly distinctive - predictions of the 
hydrodynamic model besides the global polarization along ${\bf J}$. For instance, 
in ref.~\cite{Pang:2016igs}, a connection between local vortical structures in event-by-event 
hydrodynamics and correlation of polarizations of two $\Lambda$ hyperons in transverse 
and longitudinal (along the beam line) directions has been studied. In this letter, 
we argue that in non-central heavy ion collisions, a non-zero longitudinal polarization 
of $\Lambda$ with different transverse momenta $p_T$ is a more generic effect present 
in a simple non-fluctuating hydrodynamic picture, and propose to measure it in experiment 
\cite{voloshin}. As it will be shown, this observable has several attractive features:
\begin{itemize}
\item unlike the polarization along ${\bf J}$, it is sensitive only to the 
 transverse expansion dynamics;
\item it is found not to decrease rapidly as a function of center-of-mass
energy (similar to longitudinal correlations in \cite{Pang:2016igs}) and it can 
be detected even at the LHC energy in the TeV range;
\item it survives the ``minimal vorticity" scenario of Bjorken longitudinal
boost invariance; 
\item unlike the polarization component along the angular momentum, it 
does not require the identification of the orientation \footnote{Hereafter we use 
term `orientation' in its mathematical sense, meaning discrete choice of orientation 
of a normal vector to the plane} of the reaction plane, thus greatly reducing the 
experimental labor.
\end{itemize}
The effect is dominated by the geometry of collision, therefore we do not include 
event-by-event fluctuations in this study.

\begin{figure}
\includegraphics[width=0.49\textwidth]{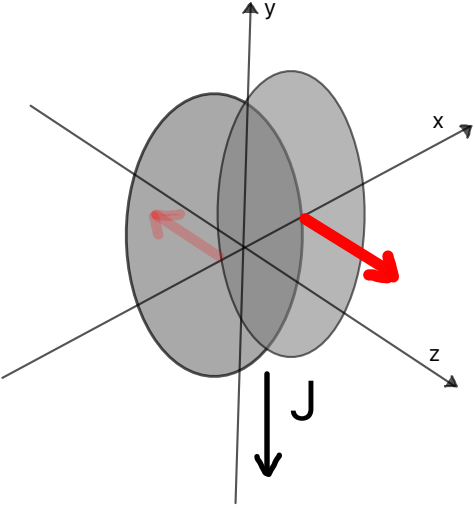}
\caption{A sketch of peripheral relativistic heavy-ion collisions. The system
is symmetric for a rotation by 180$^o$ around the angular momentum and reflection
with respect to the reaction plane $xz$. The axes $x,y,z$ are the reference frame 
for this work.}
\label{collisions}
\end{figure}
{\em Symmetries in relativistic nuclear collisions at very high energy}

In principle, (average) collisions of two identical nuclei at a finite impact 
parameter feature two initial discrete symmetries: rotation by an angle $\pi$ 
around the total angular momentum axis and reflection in the transverse plane with 
respect to the reaction plane (see fig.~\ref{collisions}). Their combination implies 
an invariance by total reflection ${\bf x} \to -{\bf x}$. 
In the high energy limit, another (continuous) symmetry becomes plausible 
and it is commonly assumed in the hydrodynamic modeling of relativistic heavy-ion
collisions, that is the invariance by Lorentz boost along the beam axis, also known 
as Bjorken longitudinal boost invariance. The straightforward consequence of the
boost invariance is that any scalar function of space-time coordinates is independent 
of the space-time rapidity $\eta = (1/2) \log [(t+z)/(t-z)]$, where $z$ is the 
Cartesian coordinate along the beam direction and $t$ the time in the center-of-mass 
frame. Also, by combining reflection with boost invariance, one readily obtains that 
the space-time rapidity component of any vector field $V$ vanishes, that is $V^\eta=0$.

An orthogonal (i.e.\ scalar product conserving) symmetry transformation for a 
general tensor field (including all special cases: scalar, vector, pseudo-vector 
etc.) can be expressed as:
\be\label{symm}
   T({\sf \Lambda}(x)) = D({\sf \Lambda}) T(x)
\ee
where ${\sf \Lambda}$ is the linear symmetry transformation and $D({\sf \Lambda})$ its 
representation matrix for the tensor field $T$. Suppose that we deal with a (tensor) 
function $\Theta$ of four-momentum which is expressed as an integral over a domain 
$\Omega$, symmetric under the transformation ${\sf \Lambda}$, of a tensor function
of $x$ and $p$:
\be\label{obsp}
 \Theta(p) = \int_\Omega \di \Omega \; T(x,p) 
\ee
with 
\be\label{symmps}
T({\sf \Lambda}(x),{\sf \Lambda}(p)) = D({\sf \Lambda}) T(x,p)
\ee
Therefore:
\bea\label{symmom}
&& \Theta({\sf \Lambda}(p)) = \int_\Omega \di \Omega \; T(x,{\sf \Lambda}(p)) \nonumber \\
&& =  \int_D \di \Omega \; D({\sf \Lambda})^{-1} T({\sf \Lambda}^{-1}(x),p)
 = D({\sf \Lambda})^{-1} \int_{{\sf \Lambda}^{-1}(\Omega)} \!\!\!\!\!\!\!\!
  \di \Omega' \; T(x',p) \nonumber \\
&& = D({\sf \Lambda})^{-1} \int_{\Omega} \; \di \Omega \; T(x',p) = 
 D({\sf \Lambda})^{-1}\Theta({\sf \Lambda}(p))
\eea
where we have used the eq.~(\ref{symmps}) and the invariance of the domain $\Omega$ under
the transformation ${\sf \Lambda}$, implying $\di\Omega'=\di\Omega$ and ${\sf \Lambda}(\Omega)
= \Omega$. We are thus led to the conclusion that, when dealing with observables in momentum
space like (\ref{obsp}) with integrand fulfilling the condition (\ref{symmps}), the 
spatial symmetries have corresponding ones in momentum space as demonstrated by the 
eq.~(\ref{symmom}). For instance, this is the case for particle momentum spectrum of fermions 
which can be written as an integral over the decoupling (or {\it particlization}) 3D 
hypersurface $\Sigma$:
$$
  \int_\Sigma \di\Sigma_\lambda p^\lambda n_F = \int_\Sigma \di\Sigma (n \cdot p) n_F 
$$
where $n$ is the unit vector perpendicular to the hypersurface $\Sigma$ and $n_F$ is
the relativistic Fermi-Dirac distribution:
$$
  n_F = \frac{1}{\e^{\beta \cdot p - \sum_j \mu_j q_j/T}+1}
$$
$\beta=(1/T)u$ being the four-temperature vector. The reason is that any integrand
function involving the scalar product of four-momentum and a symmetric vector field 
like $\beta(x)$ or $n(x)$, fulfills the eq.~(\ref{symmps}). For instance:
$$
 \beta({\sf \Lambda}(x))\cdot{\sf \Lambda}(p) = {\sf \Lambda}(\beta(x))
  \cdot{\sf \Lambda}(p) = \beta(x) \cdot p
$$
For the specific case of relativistic heavy-ion collisions and longitudinal boost 
invariance, this means that the spectrum of final particles will be invariant under 
longitudinal boost in momentum space, that is independent of the rapidity $Y=(1/2) 
\log [(E+p_z)/(E-p_z)]$, and similarly for the reflection and discrete rotation 
symmetries.
\\
{\em Polarization of emitted particles in high energy heavy-ion collisions}

\begin{figure*}
\includegraphics[width=0.98\textwidth]{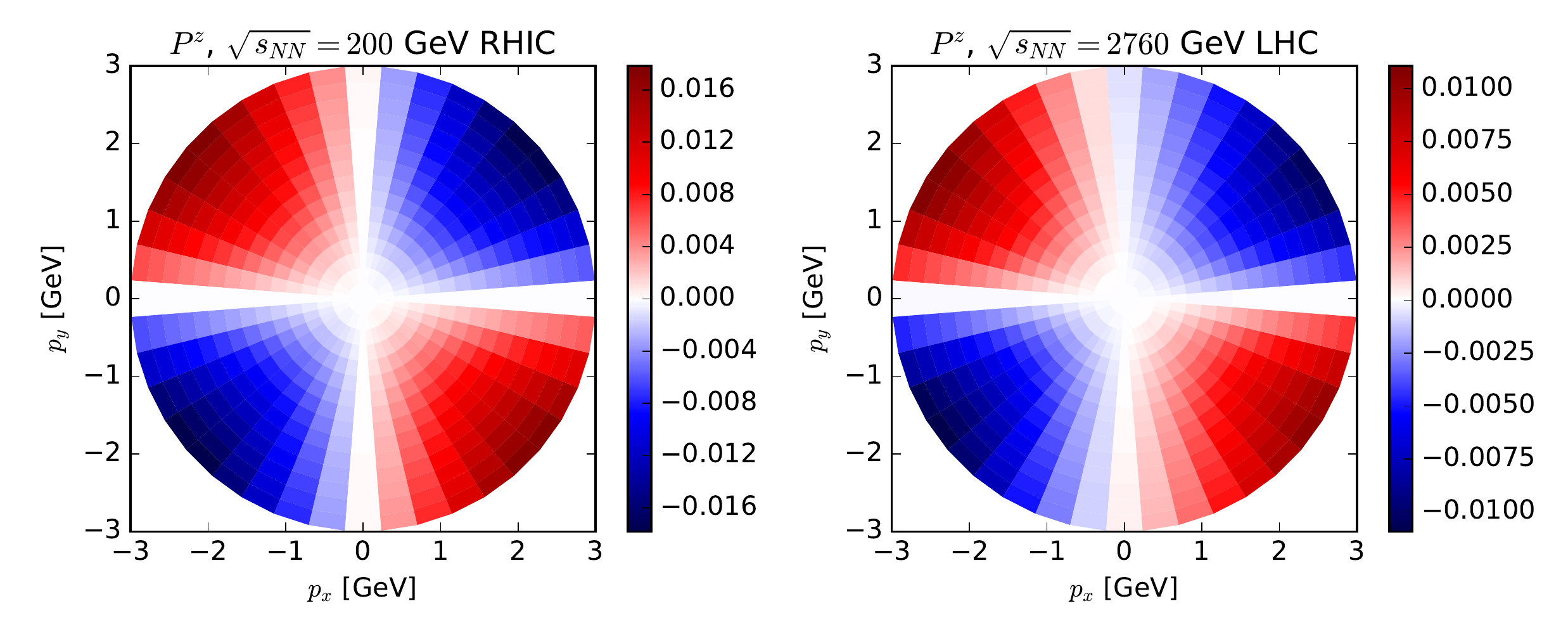}
\caption{Map of longitudinal component of polarization of midrapidity $\Lambda$ from
a hydrodynamic calculation corresponding to 20-50\% central Au-Au collisions at $\snn=200$~GeV 
(left) and 20-50\% central Pb-Pb collisions at $\snn=2760$~GeV (right).}\label{szmap}
\end{figure*}

These symmetries have remarkable consequences on the polarization of emitted particles, 
specifically on the single-particle mean spin vector $S^\mu(p)$. At the leading order, 
the mean spin vector is given by the formula \cite{Becattini:2013fla}:
\be\label{mainf}
 S^\mu(p) = - \frac{1}{8m} \epsilon^{\mu\rho\sigma\tau} p_\tau \frac{\int_\Sigma 
 \di \Sigma_\lambda p^\lambda \varpi_{\rho\sigma} n_F (1-n_F)}{\int_\Sigma \di 
 \Sigma_\lambda p^\lambda n_F} 
\ee
where $\varpi$ is the thermal vorticity, that is:
\be\label{thvort}
 \varpi_{\mu\nu} \equiv \frac{1}{2}\left(\partial_\nu \beta_\mu - \partial_\mu \beta_\nu 
 \right)
\ee
and $\Sigma$ is the decoupling hypersurface (see also ref.~\cite{Fang:2016vpj, Florkowski:2017ruc}). 
It is important to stress that in the 
eq.~(\ref{mainf}), the Cartesian coordinates in the integrand are understood as they 
are the only ones making sense of an integral of a vector field. The $S(p)$ is 
a pseudo-vector in momentum space, so, unlike polar vectors, the component along 
the beam line $S^z$ (henceforth defined as {\it longitudinal}) at $Y=0$ can be 
non-vanishing and must feature a quadrupole pattern in the transverse momentum plane 
like that shown in fig.~\ref{szmap}. Particularly, the rotation-reflection symmetries imply 
that $S^z$ has a Fourier decomposition involving only the sine of even multiples 
of the azimuthal angle $\varphi$:
\be\label{szfour}
 S^z({\bf p}_T,Y=0) = \frac{1}{2} \sum_{k=1}^\infty f_{2k}(p_T) \sin 2 k \varphi
\ee
A nice feature of the eq.~(\ref{szfour}) is that the sign of $S^z$ in the transverse
momentum plane does not depend on the reaction plane orientation, 
for a fixed handedness of the reference frame. The basically quadrupole pattern
of the longitudinal component of the spin vector had been observed in numerical 
hydrodynamic calculations \cite{Becattini:2015ska,Karpenko:2016jyx}, with a remarkable 
feature that $S^z$ has an absolute magnitude larger than those of the transverse 
components and we will delve into this feature later on.

Longitudinal boost invariance has further consequences for the spin vector in
momentum space. Because of the boost and reflection invariance, a vector field at 
$z=0$ must have vanishing Cartesian longitudinal component $V^z=0$. Likewise, the
Cartesian longitudinal component of any vector field in momentum space must be 
vanishing at midrapidity $Y=0$. This implies that the transverse components of 
a pseudo-vector field in momentum space must be vanishing, that is:
$$
 S^x ({\bf p}_T,Y=0) = S^y ({\bf p}_T,Y=0) = 0
$$

The current experimental evidence seems to bear out the asymptotic longitudinal boost
invariance scenario insofar as $S^y$, that is the component perpendicular to the reaction 
plane, is found to steadily decrease as center-of-mass energy increases. This has been 
observed by the experiment STAR \cite{STAR:2017ckg} and confirmed by a null result 
from ALICE \cite{alice} at $\snn = 2.76$ TeV. These observations are in agreement with 
numerical calculation of the polarization of $\Lambda$ hyperons carried out in 
refs.~\cite{Becattini:2015ska,Karpenko:2016jyx} with hydrodynamic models asymptotically 
fulfilling longitudinal boost invariance. In fact, according to these calculations the 
longitudinal component of the mean spin vector $S^z$ turns out to be sizeably larger 
in magnitude than the transverse ones from $\snn \approx 60$ GeV onwards and it is thus 
reasonable to surmise that it will survive at the highest center-of-mass energy in the 
TeV range. 

Indeed, it can be shown that this component does not vanish even in the exact boost 
invariant scenario with no initial state fluctuations and that it decreases slowly 
with increasing center-of-mass energy. 
For the sake of simplicity, let us demonstrate that with an explicit calculation by assuming 
that the fluid is ideal, uncharged and that the {\em initial} transverse velocities 
$u^x,u^y$ vanish. Accumulated evidence in relativistic heavy-ion collisions indicates 
that these are reasonable approximations at very high energy. Under such assumptions, 
it is known that a particular antisymmetric tensor, the T-vorticity, 
\be
 \Omega_{\mu\nu} = \partial_\mu (T u_\nu) -\partial_\nu (T u_\mu)
\ee
vanishes at all times \cite{Becattini:2015ska,Deng:2016gyh}, as a consequence of the 
equations of motion. In this case, the thermal vorticity reduces to \cite{Becattini:2015ska}:
\be\label{thvort2}
 \varpi_{\mu\nu} = \frac{1}{T} \left( A_\mu u_\nu - A_\nu u_\mu \right)
\ee
$A$ being the four-acceleration field. This form of the thermal vorticity shows its
entirely relativistic nature, its spatial part being proportional to $({\bf a}\times{\bf v})/c^2$
in the classical units. If we now substitute the eq.~(\ref{thvort2}) in the 
eq.~(\ref{mainf}), we get:
\be\label{mainf2}
S^\mu(p) = - \frac{1}{4m} \epsilon^{\mu\rho\sigma\tau} p_\tau \frac{\int_\Sigma \di \Sigma_\lambda 
  p^\lambda A_\rho \beta_\sigma n_F (1-n_F)}{\int_\Sigma \di \Sigma_\lambda p^\lambda n_F} 
\ee
which shows that $S^z(p)$ can get contributions from the vector product of fields
and momenta in the transverse plane, where they are expected to significantly
develop even in the case of longitudinal boost invariance. The uncharged perfect fluid 
equations of motion can be written as:
$$
  A_\rho = \frac{1}{T} \nabla_\rho T = \frac{1}{T} \left(\partial_\rho T -
   u^\rho u \cdot \partial T \right)
$$   
If we plug the above acceleration expression in the eq.~(\ref{mainf3}), only
the first term with $\partial_\rho T$ gives a finite contribution as the second 
term vanishes owing to the presence of $\beta_\sigma u_\rho$ factor and the 
Levi-Civita tensor. Furthermore, since:
$$
  \frac{\partial}{\partial p^\sigma} n_F = - \beta_\sigma n_F (1-n_F)
$$
we can rewrite the eq.~(\ref{mainf2}) as:
\be\label{mainf3}
S^\mu(p) = \frac{1}{4mT} \epsilon^{\mu\rho\sigma\tau} p_\tau \frac{\int_\Sigma 
\di \Sigma_\lambda  p^\lambda \frac{\partial n_F}{\partial p^\sigma} \partial_\rho T}
{\int_\Sigma \di \Sigma_\lambda p^\lambda n_F} 
\ee
We can now integrate by parts the numerator in the above equation:
$$
 \int_\Sigma \di \Sigma_\lambda p^\lambda \frac{\partial n_F}{\partial p^\sigma} \partial_\rho T
 = \frac{\partial}{\partial p^\sigma} \int_\Sigma \di \Sigma_\lambda p^\lambda n_F 
 \partial_\rho T - \int_\Sigma \di \Sigma_\sigma n_F \partial_\rho T
 $$
Another very reasonable assumption is that the decoupling hypersurface at high
energy is described by the equation $T=T_{\rm c}$ where $T_{\rm c}$ is the QCD 
pseudo-critical temperature. This entails that the normal vector to the hypersurface 
is the gradient of temperature. 
Then the final expression of the mean spin vector is:
\be\label{mainf4}
S^\mu(p) = \frac{1}{4mT} \epsilon^{\mu\rho\sigma\tau} p_\tau 
 \frac{\frac{\partial}{\partial p^\sigma}\int_\Sigma 
\di \Sigma_\lambda  p^\lambda n_F \partial_\rho T}
{\int_\Sigma \di \Sigma_\lambda p^\lambda n_F} 
\ee
%

The longitudinal component of the mean spin vector $S^z$ thus depends on the value of 
the temperature gradient on the decoupling hypersurface and its measurement can provide
information thereupon. A simple solution of the above integral appears under 
the assumption of isochronous decoupling hypersurface, with the temperature field
only depending on the Bjorken time $\tau = \sqrt{t^2-z^2}$. In this case the 
parameters describing the hypersurface are $x,y,\eta$ with $\tau=const.$ and 
the only contribution to the numerator of the (\ref{mainf4}) arises from $\rho=0$:
$$
  \int \di \Sigma_\lambda  p^\lambda n_F \frac{\di T}{\di \tau} \cosh \eta
$$
At $Y=0$, the factor $\cosh\eta$ can be approximated with $1$ because of the 
exponential fall-off $\exp[-(m_T/T) \cosh \eta]$ involved in $n_F$, therefore:
\begin{eqnarray*}
S^z({\bf p}_T,Y=0) \hat{\bf k} &\simeq& - \frac{\di T/\di \tau}{4mT} \hat{\bf k} 
\frac{\partial}{\partial \varphi} \log \int_\Sigma \di \Sigma_\lambda p^\lambda n_F  
\end{eqnarray*}
where $\varphi$ is the transverse momentum azimuthal angle, counting from the reaction plane. In the above equation the longitudinal spin component
is a function of the spectrum alone at $Y=0$. By expanding it in Fourier series
in $\varphi$ and retaining only the elliptic flow term, one obtains:
\bea\label{sz1}
S^z({\bf p}_T,Y=0) &\simeq &  - 
\frac{\di T/\di \tau}{4mT} \frac{\partial}{\partial \varphi} 2 v_2(p_T)
\cos 2\varphi \nonumber \\
&=& \frac{\di T}{\di \tau}\frac{1}{mT} v_2(p_T) \sin 2\varphi
\eea
meaning, comparing this result to eq.~(\ref{szfour}) that in this case:
$$
   f_2(p_T) = 2 \frac{\di T}{\di \tau}\frac{1}{mT} v_2(p_T)
$$
This simple formula only applies under special assumptions with regard to the hydrodynamic 
temperature evolution, but it clearly shows the salient features of the longitudinal 
polarization at mid-rapidity as a function of transverse momentum and how it can provide 
direct information on the temperature gradient at hadronization. It also shows,
as has been mentioned - that it is driven by physical quantities related to transverse 
expansion and that it is independent of longitudinal expansion.

\begin{figure}
\includegraphics[width=0.49\textwidth]{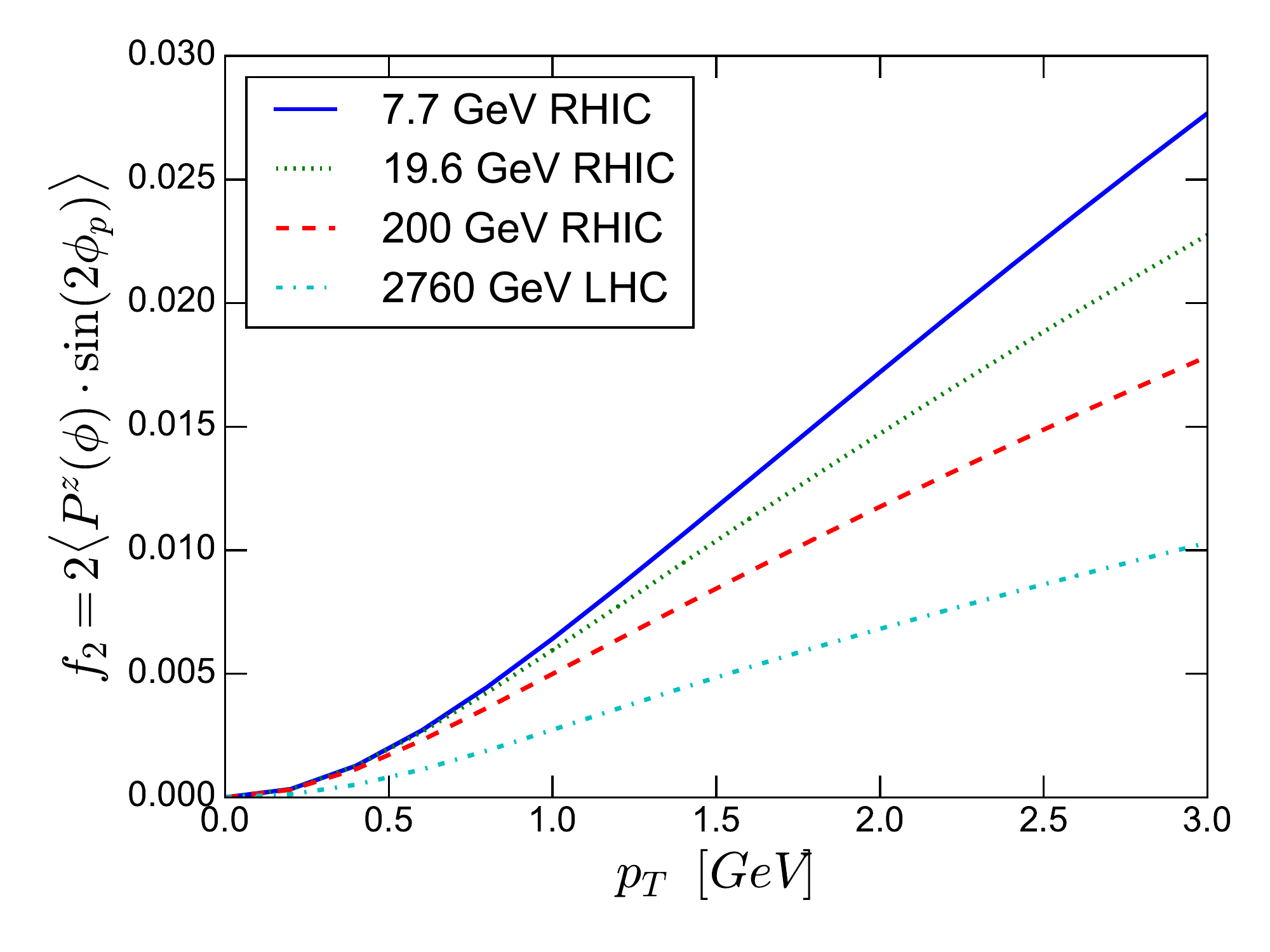}
\caption{Second harmonic of the longitudinal component of $\Lambda$ polarization $f_2$
from hydrodynamic simulations as a function of $p_T$ for different energies.}\label{fig-P2}
\end{figure}

{\em Polarization of $\Lambda$ hyperons along the beam line}\\
The above conclusion is confirmed by more realistic 3D viscous hydrodynamic 
simulations of heavy ion collisions using averaged initial state from Monte Carlo Glauber 
model with its parameters set as in \cite{Bozek:2012fw}. We have calculated the polarization 
vector ${\bf P}^*= 2 {\bf S^*}$ of primary $\Lambda$ hyperons with $Y=0$ in their rest 
frame (note that at mid-rapidity $S^{*z}=S^z$).
The resulting transverse momentum dependence of $P^{*z}$ is shown in fig.~\ref{szmap} 
for 20-50\% central Au-Au collisions at $\snn = 200$ (RHIC) and 20-50\% Pb-Pb collisions 
at $\snn = 2760$ GeV (LHC). The corresponding second harmonic coefficients $f_2$ are 
displayed in fig.~\ref{fig-P2} for 4 different collision  energies: 7.7, 19.6~GeV (calculated 
with initial state from the UrQMD cascade \cite{Karpenko:2015xea}), 200 and 2760~GeV (with 
the initial state from Monte Carlo Glauber \cite{Bozek:2012fw}). It is worth noting that, whilst 
the $P^y$ component, along the angular momentum, decreases by about a factor 10 between 
$\snn=7.7$ and 200 GeV, $f_2$ decreases by only 35\%. We also find that the mean, $p_T$ 
integrated value of $f_2$ stays around 0.2\% at all collision energies, owing to two 
compensating effects: decreasing $p_T$ differential $f_2(p_T)$ and increasing mean $p_T$ 
with increasing collision energy. 
The $P^y$ component in our calculations is produced in non-central collisions only due 
to anisotropic transverse expansion (elliptic flow), whereas in central collisions the 
initial state fluctuations dominate as shown in \cite{Pang:2016igs}. The magnitude of 
resulting correlation function (which has a $\cos(2\Delta\phi)$ shape) is similar to 
the one obtained in \cite{Pang:2016igs}.

In principle, the longitudinal polarization of $\Lambda$ hyperons can be measured 
in a similar fashion as for the component perpendicular to the reaction plane, i.e. by 
studying the distribution of $p^{*z}$, that is the longitudinal component of the 
momentum of the decay proton in the $\Lambda$ rest frame, according to the formula:
\be\label{decay}
 \frac{\di N}{\di \Omega} = \frac{1}{4\pi} \left(1+ \alpha {\bf P}^* \cdot \hat{\bf p}^*
 \right)
\ee
where $\alpha=0.642$ is the known $\Lambda$ weak decay constant. For $\Lambda$ at 
mid-rapidity, both the longitudinal polarization component and the proton $p^{*z}$ 
are the same as in the QGP frame, so the longitudinal momentum distribution 
of the decay proton is a direct probe of the mean spin vector in the QGP frame; for the 
general case, a boost must be performed, but the method is basically the same. Hence, 
at $Y=0$, the average sign of the $p^{z}$ will follow the pattern shown in fig.~\ref{szmap} 
for $S^z$, as a function of the azimuthal angle with respect to the reaction plane, with 
a leading behaviour $\sin 2\varphi$. The probability $P_s$ that the decay proton has a 
sign $s$ reads:
$$
  P_s = \frac{1}{2} + \frac{ s \alpha}{4} P^{*z}
$$  
so that the mean sign is just $(\alpha/4)P^{*z}$. 

In summary, we have shown that local thermodynamic equilibrium of the spin degrees
of freedom and the hydrodynamic model predict a global pattern of polarization 
along the beam line in relativistic heavy-ion collisions at very high energy even 
in a minimal scenario of longitudinal boost invariance, ideal fluid and no initial state fluctuations. We 
have shown that the polarization component along the beam line has a typical quadrupole 
structure of ${\bf p}_T$ dependence similar to elliptic flow, by virtue of which the 
identification of the orientation of the reaction plane is not necessary. Its measurement
is a crucial test of the hydrodynamic model and of its initial conditions and 
can provide important and unique information about the temperature gradient at the decoupling 
stage, when the Quark Gluon Plasma hadronizes around the critical temperature. Calculations 
in a realistic implementation of the hydrodynamic model indicate that its value is within 
the current reach of the experiments at RHIC and LHC energies.
\\
\begin{acknowledgments}
We are very grateful to S. Voloshin for illuminating discussions and 
clarifications. This work was partly supported by the University of Florence
grant {\it Fisica dei plasmi relativistici: teoria e applicazioni moderne}
\end{acknowledgments}



\end{document}